\documentclass{sig-alternate-2013}

\usepackage{graphicx}
\usepackage{url}
\usepackage{mathptmx}
\usepackage{amssymb}
\usepackage{amsmath}
\usepackage{subfigure}
\usepackage{hyphenat}
\usepackage{color}
\usepackage{textcomp}
\usepackage{enumitem}
\usepackage{tabularx}
\usepackage{mathtools}

\usepackage[show]{chato-notes} 

\usepackage{cmm-greek}

\newcommand{\hide}[1]{}
\newcommand{\xhdr}[1]{\vspace{1.7mm}\noindent{{\bf #1.}}}

\newcommand{\cpt}[1]{\textsc{\MakeLowercase{#1}}}

\newcommand{\rank}{\text{rank}}

\newcommand{\ie}{\textit{i.e.}}
\newcommand{\eg}{\textit{e.g.}}
\newcommand{\cf}{\textit{cf.}}

\newcommand{\vs}{\textit{vs.}}

\newcommand{\Secref}[1]{Sec.~\ref{#1}}
\newcommand{\secref}[1]{Sec.~\ref{#1}}

\newcommand{\Eqnref}[1]{Eq.~\ref{#1}}

\newcommand{\Tabref}[1]{Table~\ref{#1}}

\newcommand{\Figref}[1]{Fig.~\ref{#1}}

\newcommand{\WD}{Wikidata}
\newcommand{\WP}{Wikipedia}

\newcommand{\denselist}{ \itemsep -2pt\topsep-10pt\partopsep-10pt }

\DeclareMathAlphabet{\mathcal}{OMS}{cmsy}{m}{n}

\permission{Copyright is held by the International World Wide Web Conference Committee (IW3C2). IW3C2 reserves the right to provide a hyperlink to the author's site if the Material is used in electronic media.}
\conferenceinfo{WWW 2016,}{April 11--15, 2016, Montr\'eal, Qu\'ebec, Canada.} 
\copyrightetc{ACM \the\acmcopyr}
\crdata{978-1-4503-4143-1/16/04.\\http://dx.doi.org/10.1145/2872427.2883077}

\begin{document}%

\title{
Growing Wikipedia Across Languages via Recommendation
}

\numberofauthors{4}
\author{
\alignauthor
\hspace{-0.8cm}
Ellery Wulczyn \\
\hspace{-0.8cm}
       \affaddr{Wikimedia Foundation}\\
\hspace{-0.8cm}
       ellery@wikimedia.org
\alignauthor
\hspace{-2cm}
Robert West \\
\hspace{-2cm}
       \affaddr{Stanford University}\\
\hspace{-2cm}
       west@cs.stanford.edu
\alignauthor
\hspace{-3.5cm}
Leila Zia \\
\hspace{-3.5cm}
       \affaddr{Wikimedia Foundation}\\
\hspace{-3.5cm}
       leila@wikimedia.org
\alignauthor
\hspace{-5cm}
Jure Leskovec \\
\hspace{-5cm}
       \affaddr{Stanford University}\\
\hspace{-5cm}
        jure@cs.stanford.edu   
}

\maketitle







\begin{abstract}
The different Wikipedia language editions vary dramatically in how comprehensive they are. As a result, most language editions contain only a small fraction of the sum of information that exists across all Wikipedias. In this paper, we present an approach to filling gaps in article coverage across different Wikipedia editions. Our main contribution is an end-to-end system for recommending articles for creation that exist in one language but are missing in another. The system involves identifying missing articles, ranking the missing articles according to their importance, and recommending important missing articles to editors based on their interests. We empirically validate our models in a controlled experiment involving 12,000 French Wikipedia editors. We find that personalizing recommendations increases editor engagement by a factor of two. Moreover, recommending articles increases their chance of being created by a factor of 3.2. Finally, articles created as a result of our recommendations are of comparable quality to organically created articles. Overall, our system leads to more engaged editors and faster growth of Wikipedia with no effect on its quality.
\end{abstract}



\section{Introduction}
\label{sec:intro}

General encyclopedias are collections of information from all branches of knowledge. Wikipedia is the most prominent online encyclopedia, providing content via free access. Although the website is available in 291 languages, the amount of content in different languages differs significantly. While a dozen languages have more than one million articles, more than 80\% of Wikipedia language editions have fewer than one hundred thousand articles~\cite{meta-list-wps}. It is fair to say that one of the most important challenges for Wikipedia is increasing the coverage of content across different languages.

Overcoming this challenge is no simple task for Wikipedia volunteers. For many editors, it is difficult to find important missing articles, especially if they are newly registered and do not have years of experience with creating Wikipedia content. Wikipedians have made efforts to take stock of missing articles via collections of ``redlinks''%
\footnote{
Redlinks are hyperlinks that link from an existing article to a non-existing article that should be created.
\vspace{-12mm}
}
or tools such as ``Not in the other language''~\cite{Magnus-NIOL}. Both technologies help editors find missing articles, but leave editors with long, unranked lists of articles to choose from. Since editing Wikipedia is unpaid volunteer work, it should be easier for editors to find articles missing in their language that they would like to contribute to. One approach to helping editors in this process is to generate personalized recommendations for the creation of important missing articles in their areas of interest.

Although Wikipedia volunteers have sought to increase the content coverage in different languages, research on identifying missing content and recommending such content to editors based on their interests is scarce. Wikipedia's SuggestBot~\cite{cosley-2007} is the only end-to-end system designed for task recommendations. However,  SuggestBot focuses on recommending  existing articles that require improvement and does not consider the problem of recommending articles that do not yet exist.


Here we introduce an empirically tested end-to-end system to bridge gaps in coverage across Wikipedia language editions. Our system has several steps: First, we harness the Wikipedia knowledge graph to identify articles that exist in a source language but not in a target language. We then rank these missing articles by importance. We do so by accurately predicting the potential future page view count of the missing articles. Finally, we recommend missing articles to editors in the target language based on their interests. In particular, we find an optimal matching between editors and missing articles, ensuring that each article is recommended only once, that editors receive multiple recommendations to choose from, and that articles are recommended to the the most interested editors.

We validated our system by conducting a randomized experiment, in which we sent article creation recommendations to 12,000 French Wikipedia editors. We find that our method of personalizing recommendations doubles the rate of editor engagement. More important, our recommendation system increased the baseline article creation rate by a factor of 3.2. Also, articles created via our recommendations are of comparable quality to organically created articles. We conclude that our system can lead to more engaged editors and faster growth of Wikipedia with no effect on its quality.

The rest of this paper is organized as follows. In \Secref{sec:overview} we present the system for identifying, ranking, and recommending missing articles to Wikipedia editors. In \Secref{sec:evaluation} we describe how we evaluate each of the three system components. In \Secref{sec:experiment} we discuss the details of the large scale email experiment in French Wikipedia. We discuss some of the opportunities and challenges of this work and some future directions in \Secref{sec:discussion} and share some concluding remarks in \Secref{sec:conclusion}.





\vspace{0.2cm}

\section{System for recommending missing articles}
\label{sec:overview}

We assume we are given a language pair consisting of a \emph{source language} $S$ and a \emph{target language} $T$. Our goal is to support the creation of important articles missing in $T$ but existing in $S$.

\begin{figure}[t]
\centering
\includegraphics[width=\columnwidth]{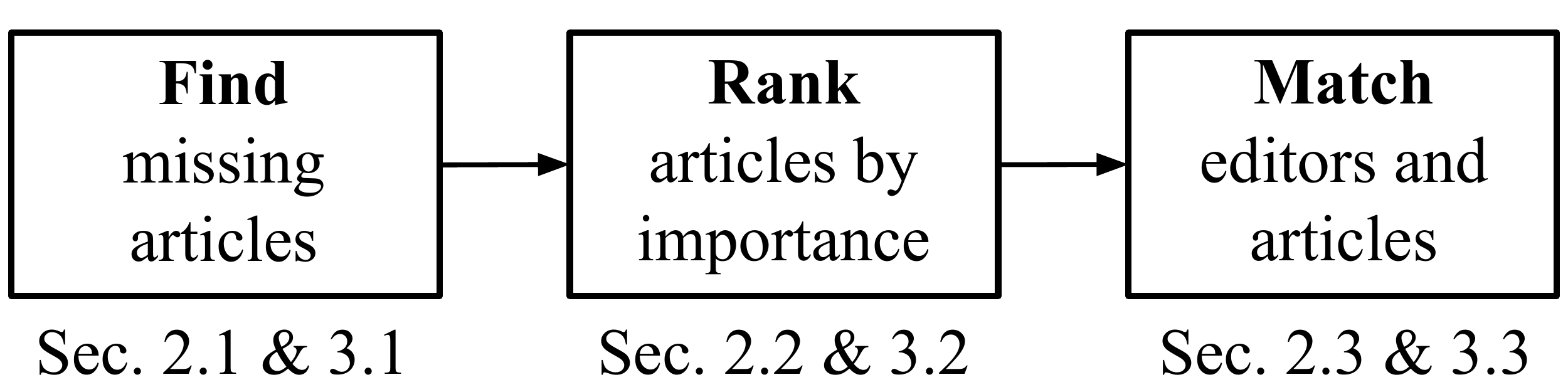}
\caption{System overview. \Secref{sec:missing}, \ref{sec:ranking}, \ref{sec:matching} describe the components in detail; we evaluate them in \Secref{subsec:missing_eval}, \ref{subsec:ranking}, \ref{subsec:matching}.}
\label{fig:system_overview}
\end{figure}

Our system for addressing this task consists of three distinct stages (\Figref{fig:system_overview}).
First, we find articles missing from the target language but existing in the source language.
Second, we rank the set of articles missing in the target language by importance, by building a machine\hyp learned model that predicts the number of views an article would receive in the target language if it existed.
Third, we match missing articles to well\hyp suited editors to create those articles, based on similarities between the content of the missing articles and of the articles previously edited by the editors.
The steps are explained in more details in the following sections.

\subsection{Finding missing articles}
\label{sec:missing}

For any pair of languages $(S, T)$, we want to find the set of articles in the \emph{source language} $S$ that have no corresponding article in the \emph{target language} $T$. We solve this task by leveraging the \emph{\WD} knowledge base \cite{wikidata}, which defines a mapping between language\hyp independent concepts and language\hyp specific Wikipedia articles. For example, the abstract concept Q133212 in Wikidata maps to 62 language\hyp specific \WP{} articles about tumors (such as the \cpt{Tumor} article in German \WP{}, \cpt{Tumeur} in French \WP{} or \cpt{Novotvorina} in Croatian \WP{}). We refer to these language\hyp independent concepts as \textit{\WD{} concepts}.

This mapping induces a clustering of the \WP{} articles from all languages, such that each cluster contains all articles about the same concept in the different languages.
Therefore, a simple approach to finding articles that are present in $S$ but missing in $T$ would be to consider those concepts whose cluster contains an article in $S$ but none in $T$. We could, for example, assume that Estonian Wikipedia has no coverage of the  \cpt{tumor} concept because the corresponding Wikidata concept Q133212 has no link to the Estonian language.

\begin{figure}[t]
\centering
\includegraphics[width=8.5cm]{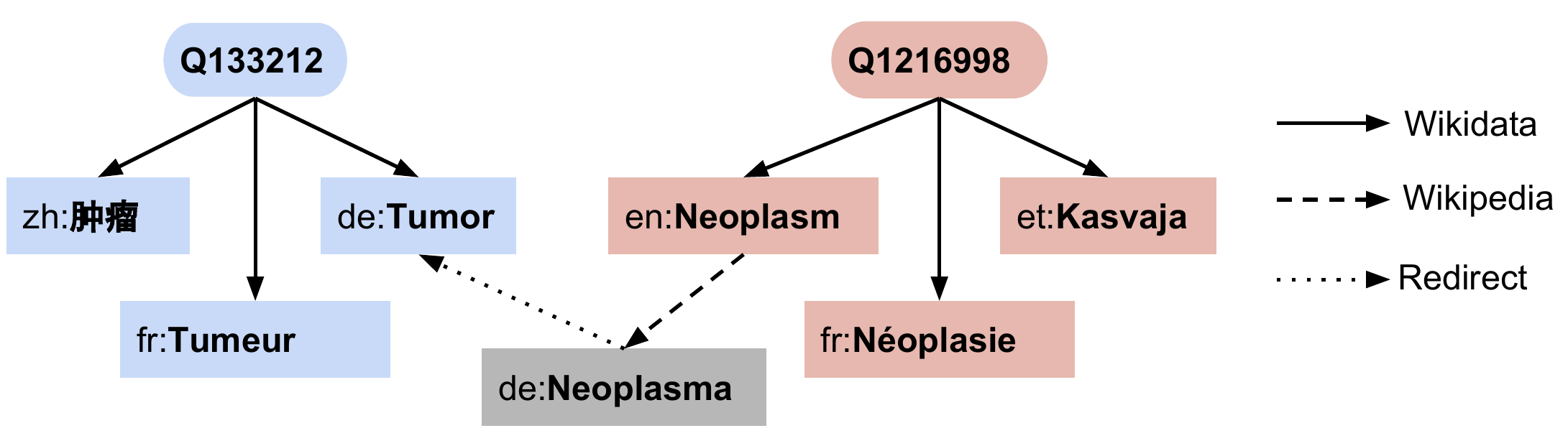}
\caption{
Language\hyp independent \WD{} concepts (oval) linking to language\hyp specific \WP{} articles (rectangular). Clusters are merged via redirects and inter\hyp language links hand\hyp coded by \WP{} editors.
}
\label{fig:neoplasm}
\end{figure}

A complicating factor is that distinct \WD{} concepts may correspond to nearly identical real\hyp world concepts, but every \WD{} concept can link to only one article per language.
For example, there are separate \WD{} concepts for \cpt{Neoplasm} and \cpt{Tumor}. The English and German Wikipedias have decided that these concepts are similar enough that each only covers one of them: German covers \cpt{Tumor}, while English covers \cpt{Neoplasm}, so the simple approach described above would consider the \cpt{Neoplasm} article to be missing in German---something we want to avoid, since the topic is already covered in the \cpt{Tumor} article.

In order to solve this problem, we need a way to partition the set of Wikidata concepts into groups of near\hyp synonyms. Once we have such a partitioning, we may define a concept $c$ to be missing in language $T$ if $c$'s group contains no article in $T$.

In order to group Wikidata concepts that are semantically nearly identical, we leverage two signals.
First, we extract inter\hyp language links which \WP{} editors use to override the mapping specified by \WD{} and to directly link articles across languages (\eg, in \Figref{fig:neoplasm}, English \cpt{Neoplasm} is linked via an inter\hyp language link to German \cpt{Neoplasma}). We only consider inter\hyp language links added between $S$ and $T$ since using links from all languages has been shown to lead to large clusters of non\hyp synonymous concepts \cite{Omnipedia}.
Second, we extract intra-language redirects. Editors have the ability to create redirects pointing to other articles in the same language (\eg, as shown in \Figref{fig:neoplasm}, German \WP{} contains a redirect from \cpt{Neoplasma} to \cpt{Tumor}).
We use these two additional types of links---inter\hyp language links and redirects---to merge the original concept clusters defined by \WD, as illustrated by \Figref{fig:neoplasm}, where articles in the red and blue cluster are merged under the same concept by virtue of these links.


In a nutshell, we find missing articles by inspecting a graph whose nodes are language\hyp independent Wikidata concepts and language\hyp specific articles, and whose edges are \WD's concept\hyp to\hyp article links together with the two additional kinds of link just described.
Given this graph, we define that a concept $c$
is missing in language $T$ if $c$'s weakly connected component contains no article in $T$.

\subsection{Ranking missing articles}
\label{sec:ranking}


Not all missing articles should be created in all languages:
some may not be of encyclopedic value in the cultural context of the target language.
In the most extreme case, such articles might be deleted shortly after being created.
We do not want to encourage the creation of such articles, but instead want to focus on articles that would fill an important knowledge gap.

A first idea would be to use the curated list of 10,000 articles every Wikipedia should have \cite{articles_every_wikipedia_should_have}.
This list provides a set of articles that are guaranteed to fill an important knowledge gap in any Wikipedia in which they are missing.
However, it is not the case that, conversely, all Wikipedias would be complete if they contained all of these articles.
For instance, an article on \cpt{Picada}, an essential aspect of Catalan cuisine, may be
crucial for
the Catalan and Spanish Wikipedias, but not
for
Hindi Wikipedia.
Hence, instead of trying to develop a more exhaustive global ranking and prioritizing the
creation of missing content according to this global ranking, we build a separate ranking for each language pair.


\xhdr{Ranking criterion}
We intend to rank missing articles according to the following criterion:
\emph{How much would the article be read if it were to be created in the given target language?}
Since this quantity is unknown, we need to predict it from data that is already known, such as the popularity of the article in languages in which it already exists, or the topics of the article in the source language (all features are listed below).

In particular, we build a regression model for predicting the normalized rank (with respect to page view counts) of the article in question among all articles in the target language.
The normalized rank of concept $c$ in language $T$ is defined as
\begin{equation}
\label{eqn:normrank}
y_T(c) := \frac{\rank_T(c)}{|T|},
\end{equation}
where $\rank_T(c)$ is the (unnormalized) rank of $T$'s article about concept $c$ when all articles in language $T$ are sorted in increasing order of page view counts. We considered page views received by each article in our dataset over the period of six months prior to the data collection point. Page view data was obtained via the raw HTTP request logs collected by the Wikimedia Foundation. Requests from clients whose user\hyp agent string reveals them as bots were excluded~\cite{ua_parser}.

For a given source--target pair $(S,T)$, the model is trained on concepts that exist in both languages, and applied on concepts that exist in $S$ but not in $T$. We experiment with several regression techniques (\Secref{subsec:ranking}), finding that random forests \cite{breiman2001random} perform best.


\xhdr{Features}
Finally, we describe the features used in the regression model for source language $S$ and target language $T$.
Here, $c_L$ is the article about concept $c$ in language $L$.
Information about $c_T$ is not available at test time, so it is also excluded during training.

\emph{Wikidata count:}
The more Wikipedias cover $c$, the more important it is likely to be in $T$.
Hence we include the number of Wikipedias having an article about $c$.

\emph{Page views:}
If $c$ is popular in other languages it is also likely to be popular in $T$.
Thus these features specify the number of page views the articles corresponding to $c$ have received over the last six months in the top 50 language versions of \WP{}.
Since some languages might be better predictors for $T$ than others, we include page view counts for each of the 50 languages as a separate feature.
(If the article does not exist in a language, the respective count is set to zero.) In addition to the raw number of page views, we also include the logarithm as well as the normalized page view rank (\Eqnref{eqn:normrank}).

\emph{Geo page views:}
If $c_S$ is popular in certain countries (presumably those where $T$ is spoken), we expect $c_T$ to be popular as well.
Hence these features specify the number of page views $c_S$ has received from each country.

\emph{Source\hyp article length:}
If $c_S$ contains substantial content we expect $c$ to be important in $T$.
Hence we consider the length of $c_S$ (measured in terms of bytes in wiki markup). 

\emph{Quality and importance classes:}
If $c_S$ is considered of high quality or importance, we expect $c$ to be an important subject in general.
To capture this, we use two signals.
First, several Wikipedias
classify articles in terms of quality as
`stub',  `good article', or `featured article' based on editor review \cite{quality_classes}.
Second, members of WikiProjects, groups of contributors who want to work together as a team to improve a specific topic area of Wikipedia, assign importance classes to articles to indicate how important the article is to their topical area \cite{importance_classes}.
We compute the maximum importance class that $c_S$ has been given by any WikiProject.
Quality and importance class labels are coded as indicator variables. 
 
\emph{Edit activity:}
The more editors have worked on $c_S$, the more important we expect $c$ to be for $T$ as well.
Hence we consider the number of editors who have contributed to $c_S$ since it was created, as well as the number of months since the first and last times $c_S$ was edited.

\emph{Links:}
If $c_S$ is connected to many articles that also exist in $T$ then the topical area of $c$ is relevant to $T$.
Therefore this feature counts the numbers of inlinks (outlinks) that $c_S$ has from (to) articles that exist in $T$.
We also include the total indegree and outdegree of $c_S$.

\emph{Topics:}
Some topics are more relevant to a given language than others.
To be able to model this fact, we build a topic model over all articles in $S$ via Latent Dirichlet Allocation (LDA) \cite{Blei:2003:LDA:944919.944937} and include the topic vector of $c_S$ as a feature.
We use the LDA implementation included in the \textit{gensim} library \cite{rehurek_lrec}, set the number of topics to 400, and normalize all topic vectors to unit length.



\subsection{Matching editors to articles}
\label{sec:matching}

Our high\hyp level objective is to encourage editors to create important articles.
We have already described how to find important missing articles (\Secref{sec:missing}, \ref{sec:ranking}).
What is needed next is to find the best editors to create those articles.
We hypothesize that editors are more likely to create articles that fall into their area of interest, and therefore we need a way of capturing how interested an editor is in creating a given article (\Secref{subsec:interest}).
Finally, in order to make the most effective recommendations, we need a way to combine the inherent importance of an article with an editor's interest in creating that article (\Secref{subsec:integration}), and to match editors with articles based on the resulting scores (\Secref{subsec:lp}).


\subsubsection{Editor interest modeling}
\label{subsec:interest}

For an editor $e$ and concept $c$ we want to score how closely $c$ matches the topics of interest to $e$. Later we will use these interest scores to match editors with missing articles. 

\begin{figure}[t]
\centering
\includegraphics[width=8cm]{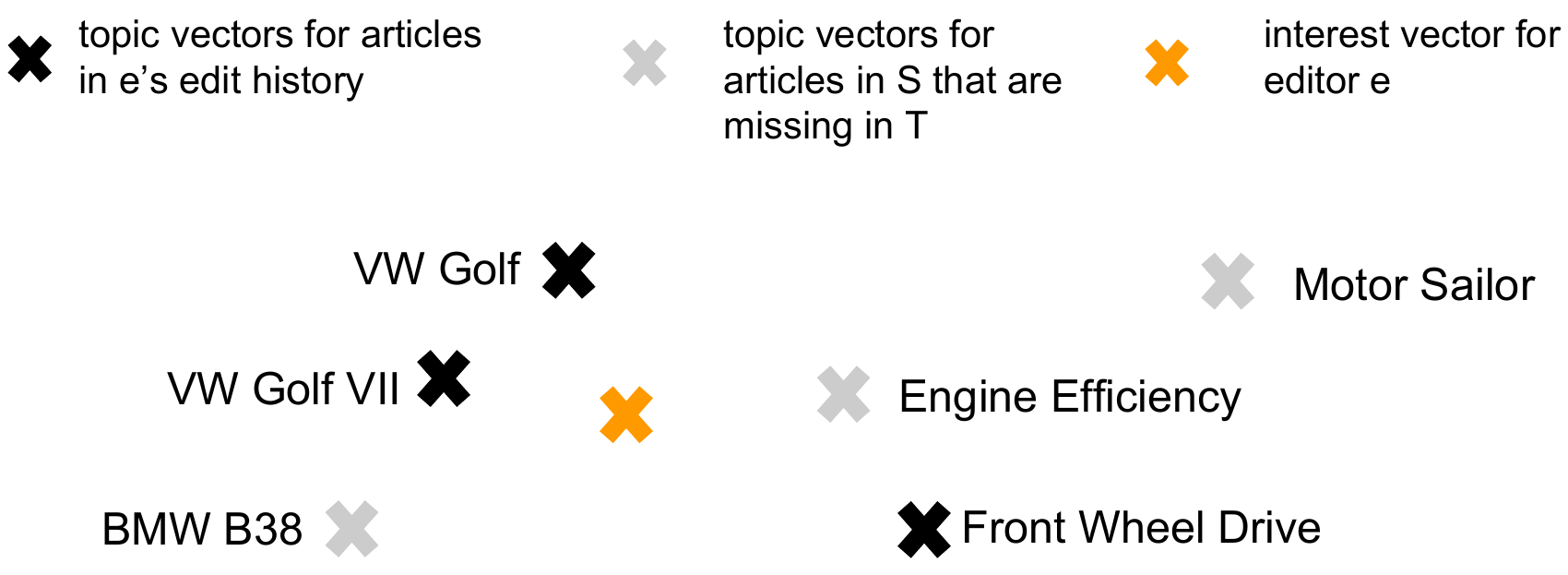}
\caption{Embedding of an editor's edit history into the topic vector space.}
\label{fig:embedding}
\end{figure}

We quantify $e$'s interest in creating an article about $c$ via the similarity of $c$ to the articles $e$ has previously edited (\ie, $e$'s \emph{edit history}). This idea is illustrated schematically in \Figref{fig:embedding}, where the black crosses represent the articles $e$ has edited, summarized in a single point by the orange cross. Gray dots stand for missing articles that could potentially be recommended to $e$.
The closer two points, the more similar the concepts they represent, and the closer a missing article (gray cross) is to $e$'s summarized edit history (orange cross), the better a recommendation it is.
Operationalizing this idea poses the challenges of
(1)~representing articles in a vector space and
(2)~aggregating an entire edit history into a single point in that vector space.

\xhdr{Vector-space representation of concepts}
First, to embed concepts in vector space, we represent the concept $c$ by the LDA topic vector of $c_S$ (\cf\ \Secref{sec:ranking}).
We can include contributions $e$ made in languages other than $S$ if the edited articles have a corresponding article in $S$. In this case, we use the topic vector of $c_S$ to represent the edited article.
The distance between two concepts is measured as the Euclidean\footnote{Since vectors are normalized, cosine distance and Euclidean distance are equivalent for our purposes.} distance between their normalized topic vectors.

\xhdr{Aggregating edit histories into interest vectors}
Second, to summarize edit histories as interest vectors, we proceed as follows.
For each revision made by $e$, we compute the number of bytes that were added to the article. We then compute the total number of bytes $e$ has added to an article over the course of all the revisions to that article. Revisions that remove bytes are not included.
This way, each concept appears at most once in each edit history.
We consider three different methods of summarizing an edit history into a single vector
in the LDA topic space, which we refer to as the editor's \emph{interest vector}
(all interest vectors are normalized to unit length):

\begin{enumerate}
\denselist
\item \textbf{Average.} The interest vector for editor $e$ is computed as the mean of the topic vectors of all articles in $e$'s edit history.
\item \textbf{Weighted average.} As above, with the difference that each concept $c$ is weighted by the logarithm of the number of bytes $e$ has added to $c$.
\item \textbf{Weighted medoid.} The interest vector is defined as the topic vector of the article from $e$'s edit history that minimizes the weighted sum of distances between it and all other topic vectors from the edit history. Weights are computed as above.
\end{enumerate}


\xhdr{History size}
When computing an interest vector, we may not want to include all articles from the edit history. For instance, some editors have worked on thousands of articles, which leads to unnecessarily long interest vector computation times and very dense interest vectors when using averaging.
An editor's interests may also evolve over time, so including articles edited a long time ago adds noise to the signal of what topics the editor is interested in now.
For these reasons, we introduce \emph{history size} as a tunable parameter, specifying the number of most recently edited articles considered for computing interest vectors.

\subsubsection{Integrating importance and interest}
\label{subsec:integration}

We aim to recommend a missing article $c_T$ to an editor $e$ if
(1)~concept $c$ is important in the target language $T$ and
(2)~it is relevant to $e$'s interests.
Above, we have proposed methods for quantifying these two aspects, but in order to make effective recommendations, we need to somehow integrate them.

A simple way of doing so is to first rank all articles missing from language $T$ by importance (\Secref{sec:missing}), then discard all but the $K$ most important ones (where $K$ is a parameter), and finally score the relevance of each remaining concept $c$ for editor $e$ by computing the distance of $c$'s topic vector and $e$'s interest vector (\Secref{subsec:interest}).

A slightly more complex approach would be to keep all missing articles on the table and compute a combined score that integrates the two separate scores for article importance and editor--article interest, \eg, in a weighted sum or product.

Both approaches have one parameter: the number $K$ of most important articles, or the weight for trading off article importance and editor--article interest.
Since we found it more straightforward to manually choose the first kind of parameter, we focus on the sequential strategy described first.

\subsubsection{Matching}
\label{subsec:lp}

When several editors simultaneously work on the same article, the danger of edit conflicts arises.
In order to prevent this from happening,
we need to ensure that, at any given time, each article is recommended to only a single editor.
Further, to avoid overwhelming editors with work, we can make only a limited number of recommendations per editor.
Finally, we want the articles that we recommend to be as relevant to the editor as possible.
Formally, these goals are simultaneously met by finding a matching between editors and articles that maximizes the average interest score between editors and assigned articles, under the constraints that each article is assigned to a unique editor and each editor is assigned a small number $k$ of unique articles.

We formulate this matching problem as a linear program \cite{edmonds1965maximum} and solve it using standard optimization software.

In practice, we find that a simple greedy heuristic algorithm gives results that are as good as
the optimal solutions obtained from the matching algorithm.
The heuristic algorithm iterates $k$ times over the set of editors for whom we are generating recommendations and assigns them the article in which they are interested most and which has not yet been assigned.





\section{Offline Evaluation}
\label{sec:evaluation}

Before evaluating our system based on deploying the complete pipeline in a live experiment (\secref{sec:experiment}), we evaluate each component offline, using English as the source language $S$, and French as the target language $T$.

\subsection{Finding missing articles}
\label{subsec:missing_eval}

Here we assess the quality of the procedure for detecting missing articles (\secref{sec:missing}). We do not assess recall since that requires a ground\hyp truth set of missing articles, which we do not have. Furthermore, using English as a source and French as the target, our procedure produces 3.7M missing articles. Given this large number of articles predicted to be missing, we are more concerned with precision than with recall.  Precision, on the other hand, is straightforward to evaluate by manually checking a sample of articles predicted to be missing for whether they are actually missing.

Our first approach to evaluating precision was to sample 100 articles uniformly at random from among the 300K most important ones (according to our classifier from \Secref{sec:ranking}). This gives 99\% precision: only one of 100 actually exists. The English \cpt{Ectosymbiosis} was labeled as missing in French, although French \cpt{Ectosymbiose} exists---it just has not been linked to \WD\ yet. Since precision might not be as high for the most important missing articles, we ran a second evaluation. Instead of taking sample test cases at random, we took a sample stratified by predicted importance of the missing article. We manually checked 100 cases in total, corresponding to ranks 1--20, 101--120, 1,001--1,020, 10,001--10,020, and 100,001--100,020 in the ranking produced by our importance model (\Secref{sec:ranking}). 

\begin{table}[t]
\centering
\begin{tabular}{rrr} \hline
\textbf{Rank}  & \textbf{Lenient precision} & \textbf{Strict precision} \\ \hline
1 &  0.85 \hspace{0.1cm} [0.64, 0.95] &  0.55 \hspace{0.1cm}  [0.34, 0.74] \\
101 & 0.90 \hspace{0.1cm} [0.70, 0.97] & 0.55  \hspace{0.1cm} [0.34, 0.74] \\
1,001 & 0.95 \hspace{0.1cm} [0.76, 0.99] & 0.90 \hspace{0.1cm} [0.70, 0.97] \\
10,001 &  1.00 \hspace{0.1cm} [0.84, 1.00] & 0.95 \hspace{0.1cm} [0.76, 0.99] \\
100,001  &  1.00 \hspace{0.1cm} [0.84, 1.00] & 0.95 \hspace{0.1cm} [0.76, 0.99] \\
\hline
\end{tabular}
\caption{
Empirical values with 95\% credible intervals for precision of missing\hyp article detector (\Secref{sec:missing}).
Rows are predicted importance levels of missing articles; for definition (also of two kinds of precision), \cf\ \Secref{subsec:missing_eval}.
}
\label{table:prec_missing_detector}
\end{table}

In addition to the unambiguous errors arising from missing \WD{} and inter\hyp language links, we observed two types of cases that are more difficult to evaluate. First, one language might spread content over several articles, while the other gathers it in one. As an example,  French covers the Roman hero \cpt{Hercule} in the article about the Greek version \cpt{H\'eracl\`es}, while English has separate articles for \cpt{Hercules} and \cpt{Heracles}.  As a consequence, the English article \cpt{Hercules} is labeled as missing in French. Whether \cpt{Hercule} deserves his own article in French, too, cannot be answered definitively. Second, a concept might currently be covered not in a separate article but as a section in another article. Again, whether the concept deserves a more elaborate discussion in an article of its own, is a subjective decision that has to be made by the human editor. Since it is hard to label such borderline cases unequivocally, we compute two types of precisions. If we label these cases as errors, we obtain \emph{strict precision;} otherwise we obtain \emph{lenient precision}.

Precision results are summarized in \Tabref{table:prec_missing_detector}. While we overall achieve good performance (lenient precision is at least 85\% across importance levels), the results confirm our expectation that detecting missing articles is harder for more prominent concepts (strict precision is only 55\% for the two highest importance levels). We conclude that it is important to ask editors to whom we make suggestions to double\hyp check whether the article is really missing in the target language. 
 Since the effort of this manual check is small compared to the effort of creating the article, this is a reasonable requirement.


\subsection{Ranking missing articles}
\label{subsec:ranking}


In this section we discuss the performance of our method for ranking missing articles by importance. Recall that here importance is measured in terms of the number of page views received by a page once it is created.
We use English as the source, and French as the target language.

\xhdr{Evaluation metrics}
We use two evaluation metrics, Spearman's rank correlation coefficient and root mean squared error (RMSE).
The former is appropriate as we care mostly about the ranking of predictions, rather than the exact predicted values.
The latter is useful because it has a natural interpretation:
since we predict normalized page view ranks (\Secref{sec:ranking}), an RMSE of, say, 0.1 means that the average article ranks 10 percentile points higher or lower than predicted.

\xhdr{Baselines}
The simplest baseline is to always predict the same constant value. We use the mean normalized rank over all articles in the target language (\ie, 0.5) as the constant and call this the \emph{mean baseline}.

It is reasonable to assume that the newly created version of an article in the target language will not be too different in terms of page view rank, compared to the version in the source language.
Hence our second baseline (termed \emph{source\hyp language baseline}) is given by the normalized rank of the already existing source version of $c$ ($y_S(c)$ in the notation of \Eqnref{eqn:normrank}).



\xhdr{Random forests}
In order to improve upon these baselines, we experimented with several regression techniques, including linear regression, ridge regression, least-angle regression, and random forests, using implementations in \textit{scikit-learn} \cite{scikit-learn}. We found that random forests \cite{breiman2001random} gave the best performance, so we focus on them. In all experiments, the data is split into a training and a testing set.
To tune the hyperparameters of the random forest model (the number of trees in the ensemble, as well as their maximum depth), we perform cross\hyp validation on the training set.

\begin{figure}[t]
\centering
\includegraphics[width=6cm]{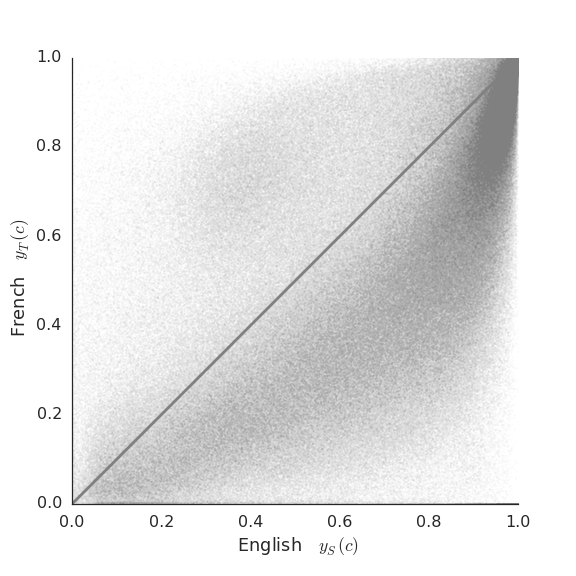}
\caption{
Scatterplot of page view ranks for articles in English \vs\ French Wikipedia.
}
\label{fig:rank_dist}
\end{figure}

\xhdr{Results}
As seen in Table \ref{table:rf}, the simple mean baseline yields an RMSE of 0.287.
The source\hyp language baseline improves on this only slightly, with an RMSE of 0.276.
\Figref{fig:rank_dist} plots the prediction of this baseline against the ground\hyp truth value (\ie, it shows a scatter plot of the normalized ranks in English and French).
We see that, while there is significant correlation, the source\hyp language rank tends to overestimate the target\hyp language rank.

\begin{table}[t]
\centering
\begin{tabular}{rll} \hline
\textbf{Model} & \textbf{RMSE} & \textbf{Spearman correlation}\\ \hline
Mean baseline & 0.287 & N/A \\ 
Source-language baseline & 0.276 & 0.673\\ 
Random forests  & 0.130 & 0.898 \\ \hline
\end{tabular}
\caption{Importance ranking test results (\Secref{subsec:ranking}).
}
\label{table:rf}
\end{table}

Table \ref{table:rf} compares the performance of the baselines with the tuned random forest regression model. The latter performs better by a large margin in terms of both RMSE (0.130 \vs\ 0.276) and Spearman correlation (0.898 \vs\ 0.673).
We conclude that leveraging additional features in a machine\hyp learned model lets us overcome the aforementioned overestimation bias inherent in the source\hyp language baseline.

\begin{figure}[t]
\centering
\includegraphics[width=8.5cm]{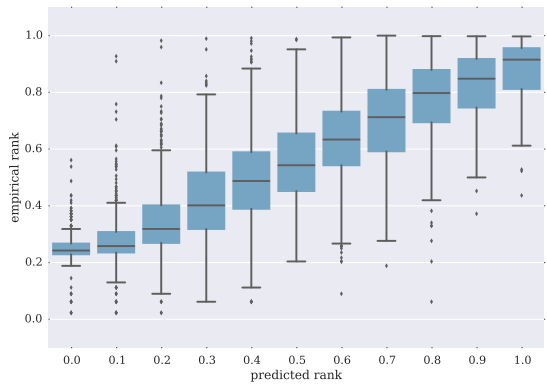}
\vspace{-1cm}
\caption{ Empirical page view ranks for newly created articles as a function of predicted ranks (for target language French). }
\label{fig:rank_vs_views}
\end{figure} 


To validate that highly ranked missing articles indeed attract more page views after being created, we performed a time\hyp based evaluation by tracking readership for the set of 5.7K English articles that were missing in French Wikipedia as of June 25, 2015, but were created by July 25, 2015. Rank predictions are made based on features of the English articles before June 25, and empirical page view ranks are computed based on traffic from July 25 through August 25. \Figref{fig:rank_vs_views} shows that the predicted ranks of these new articles correlate very well with their empirical ranks.


The RMSE between the predicted rank and the empirical rank is 0.173, which is higher than the offline validation step suggests (0.130; Table~\ref{table:rf}). This is to be expected, since empirical ranks were computed using page view counts over a single month directly after the article is created, whereas the model was trained on ranks computed using page view counts over six months for articles that may have existed for many years. Articles predicted to have a low rank that get created tend to have a higher empirical rank than predicted.
This makes sense if the creation is prompted by news events which drive both the creation and subsequent readership, and which are not anticipated by the model.



\xhdr{Feature importance}
Finally, to better understand which feature sets are important in our prediction task, we used forward stepwise feature selection. At each iteration, this method adds the feature set to the set of training features that gives the greatest gain in performance.
Feature selection is done via cross\hyp validation on the training set; the reported performance of all versions of the model is based on the held-out testing set.
For an explanation of all features, see \Secref{sec:ranking}.

\begin{table}[t]
\centering
\begin{tabular}{rlll} \hline
 &  \textbf{Feature set added} & \textbf{RMSE} & \textbf{Spearman correlation} \\ \hline
1 & Page views & 0.165 & 0.827\\ 
2 & Topics & 0.133 & 0.893 \\ 
3 & Links & 0.132 & 0.895 \\ 
4 & Geo page views & 0.131 & 0.895 \\ 
5 & Qual.\ \& import.\ classes & 0.130 & 0.898 \\ 
6 & Edit activity & 0.130 &  0.898 \\ 
7 & Source\hyp article length & 0.130 & 0.898 \\ \hline
\end{tabular}
\caption{Forward stepwise feature selection results (features explained in \Secref{sec:ranking}).
}
\label{table:feature_selection}
\end{table}

The results are listed in Table \ref{table:feature_selection}. As expected, the single strongest feature set is given by the page views the missing article gets in Wikipedias where it already exists.
Using this feature set gives an RMSE of 0.165, a significant decrease from the 0.276 achieved by the source\hyp language baseline (Table~\ref{table:rf}).
Enhancing the model by adding the LDA topic vector of the source version of the missing article results in another large drop in RMSE, down to 0.133.
Thereafter, adding the remaining feature sets affords but insignificant gains. 


\subsection{Matching editors and articles}
\label{subsec:matching}


Here we evaluate our approach to modeling editors' interests and matching editors with missing articles to be created.
Recall that we model articles as topic vectors (\Secref{subsec:interest}), and an editor $e$'s \emph{interest vector} as an aggregate of the topic vectors corresponding to the articles $e$ has edited.
To measure the predictive power of these interest vectors with regard to the edits $e$ will make next, we hold out $e$'s most recently edited article and use the remaining most recent $w$ articles  to compute $e$'s interest vector
(where $w$ is the \emph{history size} parameter, \cf\ \Secref{subsec:interest}).

Then, all articles of the source language $S$ are ranked by their distance to $e$'s interest vector, and we measure the quality of the prediction as the reciprocal rank of the held\hyp out article in the ranking. The quality of a method, then, is the the mean reciprocal rank (MRR) over a test set of editors.

\begin{figure}[t]
\centering
\includegraphics[width=8.5cm]{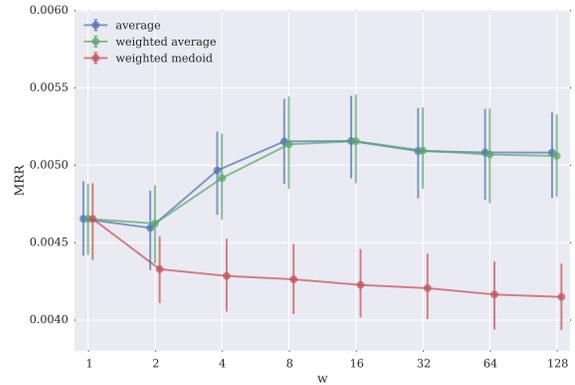}
\vspace{-1cm}
\caption{
Mean reciprocal rank of predicting the article a user edits next, based on her previous $w$ edits (logarithmic $x$-axis), with bootstrapped 95\% confidence intervals; one curve per aggregation method.
}
\label{fig:cv1}  
\end{figure}

\Figref{fig:cv1} explores how performance depends on the history size $w$ and the edit\hyp history aggregation method (\Secref{subsec:interest}), for a set of 100,000 English Wikipedia editors who have contributed to at least two articles.
We observe that the average and weighted\hyp average methods perform equally well (considerably better than weighted medoids).
Their performance increases with $w$ up to $w=16$ and then slowly decreases, indicating that it suffices to consider short edit\hyp history suffixes.
Under the optimal $w=16$ we achieve an MRR of 0.0052. That is, the harmonic mean rank of the next article an editor edits is $1/0.0052=192$. Keeping in mind that there are 5M articles in English Wikipedia, this is a good result, indicating that editors work within topics as extracted by our LDA topic model.

\section{Online Experiment}
\label{sec:experiment}

The challenge we address in this research is to boost the creation of articles purveying important knowledge.
We now put our solution to this challenge to the test in an \textit{in vivo} experiment.
We identify important missing articles, match them to appropriate editors based on their previous edits, and suggest to these editors by email that they might be interested in creating those articles.
We focus on the source\slash target pair English\slash French and, to lower the participation threshold, give contacted editors the option to use a content translation tool built by the Wikimedia Foundation \cite{cx}.

To assess the effectiveness of our system, we then ask the following research questions:
\begin{enumerate}
\denselist
\item[\textbf{RQ1}] Does recommending articles for creation increase the  \emph{rate at which they are created} compared to the rate at which articles are organically created in Wikipedia?
\item[\textbf{RQ2}] Do our targeted recommendations increase \emph{editor engagement,} compared to a scenario where we ask editors to create randomly assigned important missing articles?
\item[\textbf{RQ3}] How high is the \emph{quality of articles} created in response to our targeted recommendations?
\end{enumerate}


\subsection{Experimental design}
\label{sec:Experimental design}

In order to measure the outcomes of recommending missing articles to editors, we need a set of articles as well as a set of editors. The set of articles included in the experiment consists of the top 300K English articles missing in French  (\Secref{sec:missing}) in the importance ranking (\Secref{sec:ranking}). These articles were assigned to three groups A1, A2, and A3 (each of size 100K) by repeatedly taking the top three unassigned articles from the ranking and randomly assigning each to one of the three groups. This ensures that the articles in all three groups are of the same expected importance. There were 12,040 French editors who made an edit in the last year and displayed proficiency in English (for details, \cf\ Appendix \ref{app:editor_and_article_selection}.). These editors are suitable for receiving recommendations and were randomly assigned to treatment groups E1 and E2. All other French Wikipedia editors were assigned to the control group E3. Within E3, there are 98K editors who made an edit in the last year.

\begin{figure}[t]
\centering
\includegraphics[width=8cm]{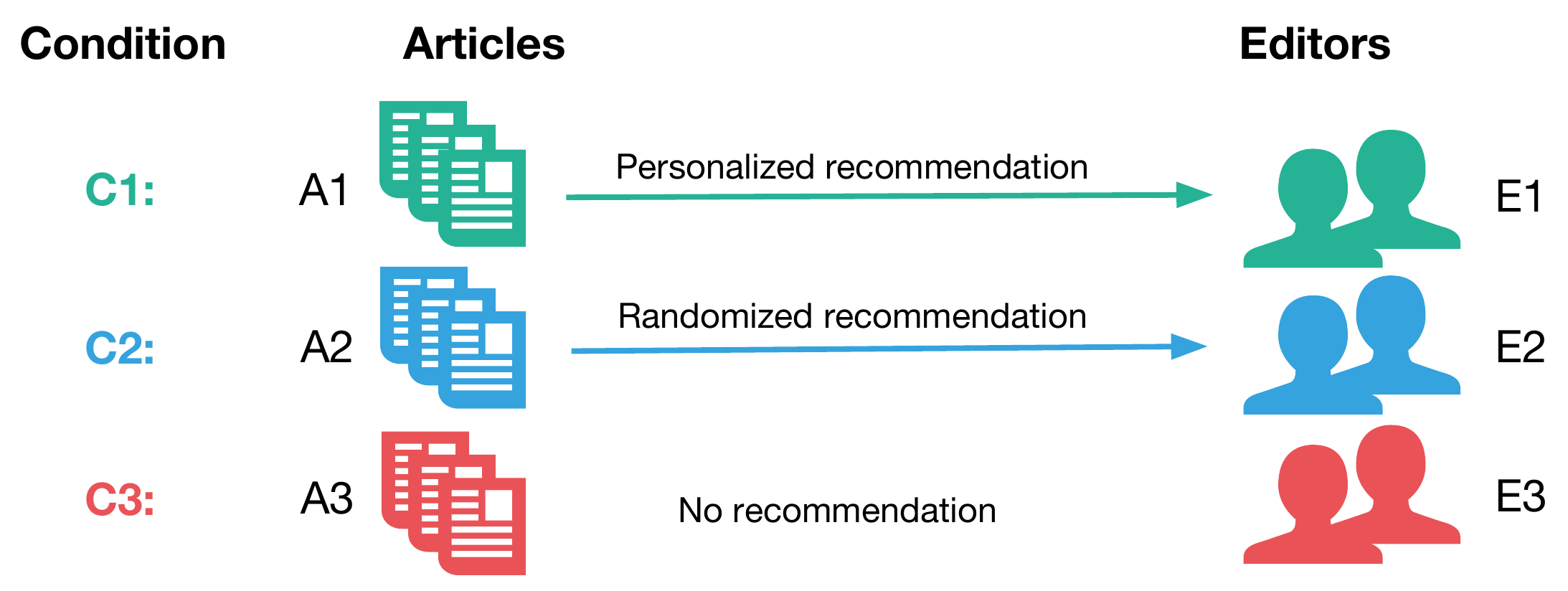}
\caption{Experimental design of user test.}
\label{fig:design}
\end{figure}

Based on these groupings, we define
three
experimental conditions (\Figref{fig:design}):


\begin{enumerate}
\denselist
\item[\textbf{C1}] \textbf{Personalized recommendation:} Editors in group E1 were sent an email recommending five articles from group A1 obtained through our interest\hyp matching method (\Secref{sec:matching}).
\item[\textbf{C2}] \textbf{Randomized recommendation:} Editors in group E2 were sent an email recommending five articles selected from group A2 at random.
(In both conditions C1 and C2, each article was assigned to at most one editor.)
\item[\textbf{C3}] \textbf{No recommendation:} Articles in group A3 were not recommended to any editor. Editors in group E3 did not receive any recommendations.
\end{enumerate}

Note that not all articles from groups A1 and A2 were recommended to editors: each group contains 100K articles, but there are only about 6K editors in each of E1 and E2; since every editor received five recommendations, only about 30K of the 100K articles in each group were recommended.

\begin{figure}[t]
\centering
\includegraphics[width=8cm]{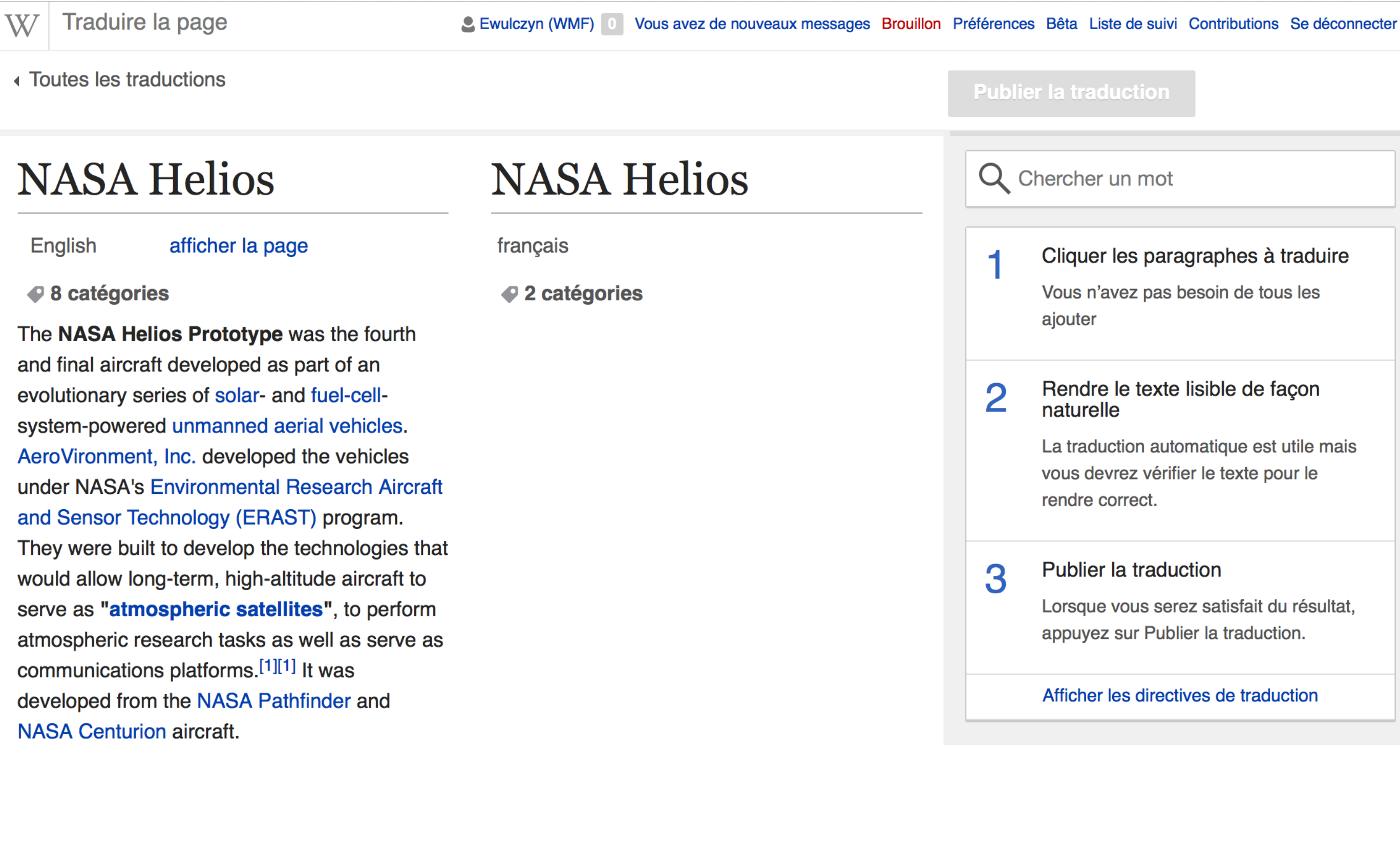}
\vspace{-0.5cm}
\caption{Screenshot of the content translation tool. The user is in the process of translating the \cpt{NASA Helios} article from English to French.}
\label{fig:cxtool2}
\end{figure}

Emails were sent to the editors in conditions C1 and C2 on June~25, 2015.
The emails were written in French and generated from the same template across conditions (\cf\ project website \cite{metapage} for exact text).
To facilitate the article\hyp creation process, each of the five recommendations was accompanied with a link to the translation tool that allows for easy section\hyp by\hyp section translation from the source to the target language (\Figref{fig:cxtool2}).


\subsection{Results}

We now evaluate the data collected through the experiment described above to answer research questions RQ1, RQ2, and RQ3.

\subsubsection{RQ1: Effect on article creation rates}
We measure the \emph{article creation rate} for each condition as the fraction of articles in the respective group that were created in the one\hyp month period after the recommendation email was sent.
(Appendix~\ref{app:counting_articles} provides methodological details on how we count the articles created in each condition.)

Comparing the article creation rates of conditions C1 and C3 will let us estimate the effect of personalized recommendation on the probability that an article is created. Further, note that the only difference between conditions C1 and C2 is that in C1 articles are assigned to editors based on our recommendation method, whereas in C2 articles are assigned to editors randomly. Therefore, by comparing the article creation rates of conditions C1 and C2, we may address the potential concern that a boost in article creation rate might be caused by the mere fact that an email recommending \emph{something} was sent, rather than by the personalized recommendations contained in the email.

\begin{table}[t]
\centering
\begin{tabular}{rrrrr} \hline
& \textbf{C1} & \textbf{C2} & \textbf{C3}   \\ \hline
Potentially created &   30,055 &    30,145 &    100,000  \\
Actually created &      316 &       177 &       322  \\
Creation rate &         1.05\% &    0.59\% &    0.32\%  \\
\hline
\end{tabular}
\caption{Article creation rates for the experimental conditions defined in \Secref{sec:Experimental design}.
}
\label{table:article_results}
\end{table}

Table~\ref{table:article_results} shows the article creation rates for all experimental conditions.
Important articles not recommended to any editor (C3) had a background probability of 0.32\% of being organically created within the month after the experiment was launched.
This probability is boosted by a factor of 3.2 (to 1.05\%) for articles that were recommended to editors based on our interest\hyp matching method (C1).
On the other hand, articles that were recommended to editors on a random, rather than a personalized, basis (C2) experienced a boost of only 1.8 (for a creation rate of 0.59\%).%
\footnote{All pairs of creation rates are statistically significantly different ($p < 10^{-7}$ in a two\hyp tailed two\hyp sample $t$-test) with highly non\hyp overlapping 95\% confidence intervals.}


A possible confound in comparing the creation rates in C1 and C2 to C3 is the possibility that our recommendation emails diverted effort from articles in C3 and that, consequently, the creation rate in C3 is an underestimate of the organic creation rate.  Although the number of editors in E1 and E2 is small (6K each) compared to the number of editors in E3 (over 98K), they differ in that editors in E1 and E2 showed proficiency in English, which might be correlated with high productivity. To address this concern, we computed the creation rate for the top 300K most important missing articles in the month prior to the experiment. We found a creation rate of 0.36\%, which is only slightly higher than the rate we observed in C3 (0.32\%). This indicates that the degree of underestimation in C3 is small. 

\begin{figure}[t]
\centering
\includegraphics[width=8cm]{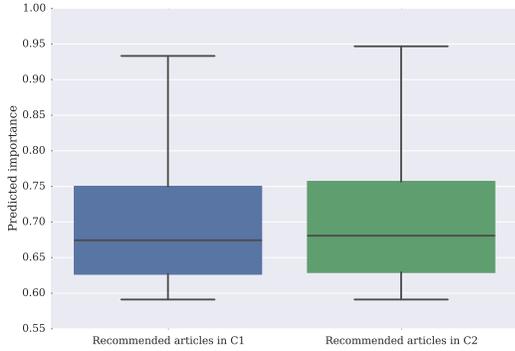}
\vspace{-0.8cm}
\caption{Box plots for predicted page view rank for articles in experimental conditions C1 and C2.
}
\label{fig:cdf}
\end{figure}

A possible confound in comparing the creation rates between C1 and C2 is that, if articles matched via personalization are also predicted to be more important, then the boost in creation rate might not stem from topical personalization but from the fact that more important articles were recommended. To investigate this possibility, we compare the predicted importance (\Secref{sec:ranking}) of the articles recommended in condition C1 and C2. \Figref{fig:cdf} shows that the two distributions are nearly identical, which implies that the boost in article creation rate is not mediated merely by a bias towards more popular articles among those recommended in C1.

We conclude that recommending articles to suitable editors based on their previously edited articles constitutes an effective way of increasing the creation rate of articles containing important knowledge.
Although some of this increase is caused merely by reminding editors by email to contribute, the quality of the specific personalized suggestions is of crucial importance.

\subsubsection{RQ2: Effect on editor engagement}
We continue our evaluation with a more editor\hyp centric analysis.
To gauge the effectiveness of our method in terms of editor engagement, we pose the following questions:
What fraction of contacted editors become active in creating a recommended article?
Are editors more likely to become active in response to an email if they receive personalized, rather than random, recommendations?

Formally, we define an \emph{active editor} as an editor who starts working on a recommended article in the translation tool (without necessarily publishing it), and a \emph{publishing editor} as one who starts and publishes a recommended article.
Given these definitions, we compute \emph{activation and publication rates} as the fractions of all editors in each group who become active or published, respectively.

\begin{table}[t]
\centering
\begin{tabular}{rrr} \hline
& \textbf{Personal (C1)} & \textbf{Random (C2)} \\ \hline
Editors contacted   & 6,011 & 6,029 \\
Active editors      & 258 & 145 \\
Publishing editors  & 137 & 69 \\
Activation rate     & 4.3\% & 2.4\% \\
Publication rate    & 2.3\% & 1.1\% \\
\hline
\end{tabular}
\caption{Effect of personalizing the recommendations.}
\label{table:user_results}
\end{table}



Table~\ref{table:user_results} compares these rates between the personalized (E1) and randomized (E2) editor groups (corresponding to experimental conditions C1 and C2), showing that about one in fifty editors in the randomized group (E2) started a recommended article in the translation tool (activation rate 2.4\%), and that half of them went on to publish the newly created article (publication rate 1.1\%).
In the personalized group (E1), on the other hand, the activation (publication) rate is 4.3\% (2.3\%); \ie, personalization boosts the activation as well as the publication rate by a factor of about two.%
\footnote{Publication and activation rates in the two conditions are statistically significantly different ($p < 10^{-5}$ in a two\hyp tailed two\hyp sample $t$-test) with highly non\hyp overlapping 95\% confidence intervals.}
This clearly shows that personalization is effective at encouraging editors to contribute new important content to Wikipedia.

\begin{figure}[t]
\centering
\includegraphics[width=8cm]{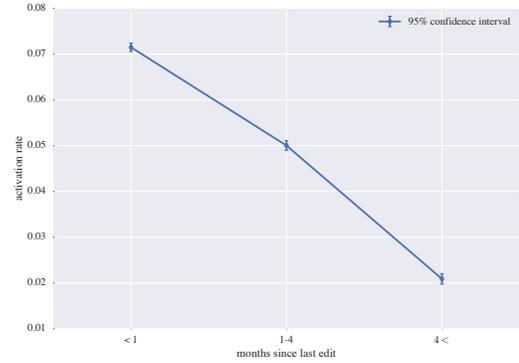}
\vspace{-0.8cm}
\caption{Editor activation rate in condition C1 as a function of months since last edit.
}
\label{fig:activity}
\end{figure}

\xhdr{Recency of activity}
The set of 12K editors who received recommendation emails included editors proficient in both English and French and having made at least one edit within the 12 months prior to the experiment (\cf\ Appendix~\ref{app:editor_and_article_selection}).
However, being active many months ago does not necessarily imply being currently interested in editing.
So, in order to obtain a more fine\hyp grained understanding of activation rates, we bucketed editors into three roughly equally sized groups based on how many months had passed between their last edit and the experiment.
\Figref{fig:activity} shows that users who were active recently are much more likely to participate, for an activation rate of 7.0\% among editors with at least one edit in the month before the experiment (compare to the overall 4.3\%; Table~\ref{table:user_results}).

\subsubsection{RQ3: Article quality}

We conclude the analysis of results by evaluating the quality of articles created with the help of our recommendations.
This is important as it is conceivable that editors might work more thoroughly when creating articles of their own accord, compared to our situation, where editors are extrinsically prompted to create new content.

\xhdr{Deletion rate}
We define the deletion rate as the percent of newly created articles deleted within 3 months of being created. 
The deletion rate for articles in C1 that were published in response to recommendations is 4.8\%, 95\% CI [2.6\%, 8.6\%], while the deletion rate for articles in C2 that were published in response to recommendations is 9.3\%, 95\% CI [4.8\%, 17.3\%]. Note that this difference is not significant ($p=0.063$). The aggregate deletion rate of articles published in response to recommendations (conditions C1 and C2) is 6.1\%, 95\% CI [3.9\%, 9.4\%]. In comparison, the overall deletion rate of articles created in French Wikipedia in the month following the experiment is vastly higher at 27.5\%, 95\% CI [26.8\%, 28.2\%] ($p<0.001$).

\begin{figure}[t]
\centering
\includegraphics[width=8cm]{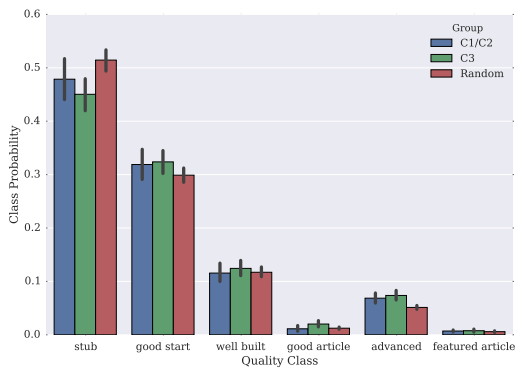}
\vspace{-0.5cm}
\caption{Aggregated quality class probabilities for articles in conditions C1 and C3, as well as for a set of 1,000 randomly selected French Wikipedia articles (with bootstrapped  95\% confidence intervals). }
\label{fig:quality}
\end{figure}

\xhdr{Automatic quality score}
We use the article quality classifier built for French Wikipedia \cite{Warncke-Wang:2013:TMM:2491055.2491063,wikiclass} to assess the quality of articles created by recommendation. Given an article, the model outputs the probability that the article belongs to each of the six quality classes used in French Wikipedia. Fig. \ref{fig:quality} shows the averaged quality class probabilities for articles created and published in response to recommendations (conditions C1 and C2) and for articles that were organically created but are of similar estimated importance (condition C3) 3 months after creation. As a baseline, we also include the distribution for a random sample of French Wikipedia articles. Articles created based on recommendations are of similar estimated quality compared to articles that were organically created and the average French Wikipedia article.

\begin{figure}[t]
\centering
\includegraphics[width=8cm]{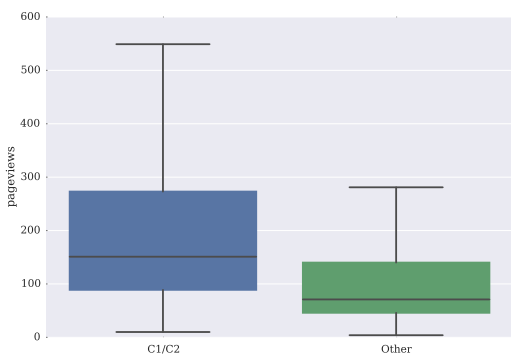}
\vspace{-0.5cm}
\caption{Box plots for the number of page views received in the first three months after creation for articles created due to a recommendation (\emph{C1/C2}) and for all other French articles created in the month after the start of the experiment (\emph{Other}). }
\label{fig:pv}
\end{figure}

\xhdr{Article popularity}
Although not directly related to article quality, we include here a brief comparison of article popularity. \Figref{fig:pv} shows the distributions over the number of page views received in the first 3 months after creation for articles created due to a recommendation and for all other French Wikipedia articles created in the month after the start of the experiment. On average, articles created based on a recommendation attract more than twice as many page views as organically created articles.

\subsubsection{Summary}

In conclusion, personalized recommendations from our system constitute an effective means of accelerating the rate at which important missing articles are created in a given target language. The chance of an article recommended via personalization being created is three times that of a similar article being created organically. Further, personalizing recommendations to the emailed editor adds significant value over sending randomly selected recommendations via email, in terms of both article creation rate and editor engagement.
Finally, the articles created in response to our targeted recommendations are less likely to be deleted than average articles, are viewed more frequently, and are of comparable quality.


\section{Discussion and Related Work}
\label{sec:discussion}

Each Wikipedia language edition, large or small, contains significant amounts of information not available in any other language~\cite{filatova2009,hecht2010}. In other words, languages form barriers preventing knowledge already available in some editions of the free encyclopedia from being accessible to speakers of other languages~\cite{adar2009,yeung2011}.

Until recently, very little was done to support cross-language content creation, with the exception of a few initiatives in specific topic areas such as the translation of medical information at the height of the Ebola crisis \cite{who15}. Multilingual contributors have been identified as playing a key role in transferring content across different language editions, particularly in smaller Wikipedias which still struggle to reach a critical mass of editors \cite{hale14}.

Part of this research uses knowledge from the rich literature on personalized task recommendation. Instead of going over that literature exhaustively, we refer the interested reader to the state of the art research on personalized task recommendation systems in crowdsourcing environments \cite{geiger-2014}.

In the rest of this section, we discuss how future research can help address some of the challenges we faced in our work.

\begin{figure}[t]
\centering
\includegraphics[width=8cm]{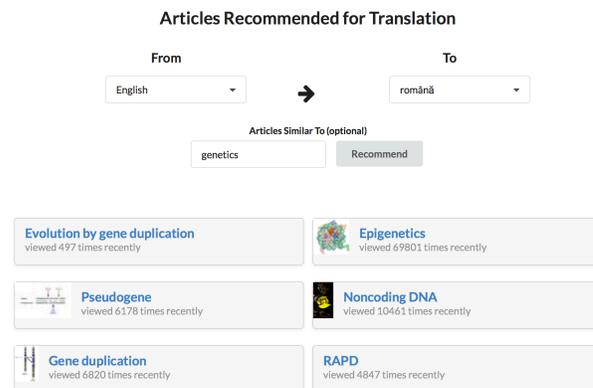}
\caption{Screenshot of the Web application for translation recommendation.}
\label{fig:app}
\end{figure}

\xhdr{Email campaigns}
Email campaigns such as the one conducted in this research are limiting in several respects. Emailing recommendations involves contacting editors who may not be interested in receiving recommendations. On the other hand, editors who enjoy the recommendations may wish to receive more of them. To address these issues, we built a Web application\footnote{http://recommend.wmflabs.org} that allows users to request missing article recommendations for several language pairs. In order to make the tool useful for new editors without an edit history and to allow existing editors to explore different interests, we prompt users for a seed article. The topic vector for the seed article is used analogously to the user's \textit{interest vector} and is used to generate personalized recommendations as described in \Secref{subsec:interest}. \Figref{fig:app} shows the relevant missing articles generated by the application for a user interested in translating articles on \cpt{Genetics} from English to Romanian.

\xhdr{Incentives to contribute}
As part of this research we did not test for the impact of different incentive mechanisms. The only incentivizing information participants received was that the recommended articles were important and missing in their language. Future research on the effect of different incentive mechanisms and how much they can increase or sustain the observed boost in the article creation rate is a promising direction.

\xhdr{The measure of importance}
In this work we rank the missing articles with respect to their predicted page views once they are created. However, it is debatable whether page views should be used as the sole measure of importance. For example, an article which is predicted to be widely read in a language may not meet the requirements for notability \cite{notability} in that language project even if the article exists in one or more other languages. This is because the notability policies and practices are sometimes different in different Wikipedia language projects. Using notability as a substitute for predicted page views has the limitation that building a good training set is hard. Although many articles have been deleted from Wikipedia due to the lack of their notability, this is not the only reason for deletion and not all articles that are still in Wikipedia are notable. Research in identifying better measures of importance for article ranking can improve the quality of the recommendations.

\xhdr{Language imperialism and translation}
An editor contacted in the experiment described recommending translations from English to French as an act of ``language imperialism''. Providing only English as a source language would imply that all concepts worth translating are contained in English Wikipedia, that only non-English Wikipedias need to be augmented by translation, and that out of all Wikipedia articles that cover a concept, only the English version should be propagated. A related concern is that different Wikipedia language editions cover the same concepts very differently \cite{hecht2010} and that fully translated articles may fail to contain important information relevant to a particular language community. A major advantage of computer-supported human translation over the current state-of-the-art in machine translation is that human translators who understand the culture of the target language can alter the source where appropriate. An interesting avenue of further research would be to compare the cultural differences expressed in the initial version as well as the revisions of translated articles with their source texts.

\xhdr{Knowledge gaps}
In this work we focused on missing articles that are available in one language but missing in another. There are multiple directions in which future research can expand this work by focusing on other types of missing content. For example, an article may exist in two languages, but one of the articles may be more complete and could be used to enhance the other. Given a method of determining such differences, our system could easily be extended to the task of recommending articles for enhancement. Alternatively, there may be information that is not available in any Wikipedia language edition, but is available on the web. The TREC KBA research track \cite{frank2012} has focused on this specific aspect, though their focus is not limited to Wikipedia. By personalizing the methodologies developed by TREC KBA research one could help identify and address more knowledge gaps in Wikipedia.

\section{Conclusion}
\label{sec:conclusion}

In this paper we developed an end-to-end system for reducing knowledge gaps in Wikipedia by recommending articles for creation. Our system involves identifying missing articles, ranking those articles according to their importance, and recommending important missing articles to editors based on their interests. We empirically validated our proposed system by running a large-scale controlled experiment involving 12K French Wikipedia editors.
We demonstrated that personalized article recommendations are an effective way of increasing the creation rate of important articles in Wikipedia. We also showed that personalized recommendations increase editor engagement and publication rates. Compared to organically created articles, articles created in response to a recommendation display lower deletion rates, more page views and comparable quality,

In summary, our paper makes contributions to the research on increasing content coverage in Wikipedia and presents a system that leads to more engaged editors and faster growth of Wikipedia with no effect on its quality. We hope that future work will build on our results to reduce the gaps of knowledge in Wikipedia.

\xhdr{Acknowledgements}
{\small
The authors would like to thank
Dario Taraborelli for fruitful discussions,
the Wikimedia Foundation Language Engineering team for technical support prior,
and the French Wikipedia community for their engagement.
This research has been supported in part by NSF
CNS-1010921,            
IIS-1149837,              
NIH BD2K,
NIH R01GM107340,
ARO MURI,
DARPA XDATA,
DARPA SIMPLEX,
Stanford Data Science Initiative,
Boeing,                  
SAP,                      
and Volkswagen.
}

\bibliographystyle{abbrv}
\bibliography{missing-articles}

\appendix

\section{Online experiment:\\Methodological details}

\subsection{Editor and article selection}
\label{app:editor_and_article_selection}

\xhdr{Editor selection}
Only editors with high proficiency in both English and French are suitable for translating from English to French. Editors can explicitly signal their proficiency level in different languages on their user pages using the \textit{Babel} template \cite{babel}.
Editors of French Wikipedia who signaled high proficiency in English were included in the experiment.
We also included editors who made an edit in both French and English Wikipedia in the 12 months before the experiment (regardless of their use of the \textit{Babel} template), assuming that these editors would be proficient in both languages.
Since the same editor can have different user names on different Wikipedias, we use the email addresses associated with user accounts to determine which English and French accounts correspond to the same editor.
We obtained a total of 12,040 editors.

\xhdr{Article selection}
We find English articles missing in French
using the method described in \Secref{sec:missing}. We excluded disambiguation pages, very short articles (less than 1,500 bytes of content) and rarely read articles (less than 1,000 page views in the last 6 months).

\subsection{Counting articles created}
\label{app:counting_articles}

\xhdr{Counting recommended articles created}
The translation tool starts logging an editor's actions after they have signed into Wikipedia, chosen a French title for the recommended missing article, and started translating their first section. This makes determining if an editor became active in response to the recommendation email easy if they used the tool.
A complicating factor is that there were 39 editors who published a recommended translation (as captured by the publicly available edit logs) but did not engage with the translation tool at all. We also consider these editors to be active.

\xhdr{Counting articles created organically}
To determine the organic creation rate of article group A3, we need to determine which newly created French articles previously existed in English, and divide by the number of articles that previously existed only in English and not in French.
We observe that, within one month of being created, nearly all new French articles (92\%) were linked to Wikidata, so to determine if a new article had an English version, we may simply check if the respective Wikidata concept had an English article associated with it.
62\% of the new French articles meet this criterion.
Further, since many newly created articles are spam and quickly deleted, we only consider articles that have persisted for at least one month.
This defines the 324 articles created organically in condition C3 (Table~\ref{table:article_results}).


\end{document}